\documentclass[%
 reprint,
 amsmath,amssymb,
 aps,
 prl,
 floatfix,
]{revtex4-1}

\usepackage{graphicx}
\usepackage{dcolumn}
\usepackage{bm}

\usepackage{url}
\usepackage{subfig}
\usepackage{xcolor}
\usepackage{tikz}
\usetikzlibrary{arrows}
\usepackage[normalem]{ulem}
\usepackage{soul}

\newcommand{\change}[1]{\textcolor{black}{#1}}
\newcommand{\materialsN}{\change{26,739}}
\begin{document}

\preprint{APS/123-QED}

\title{Online Search Tool for Graphical Patterns in Electronic Band Structures}

\author{Stanislav~S.~Borysov$^{1}$}
\email{borysov@kth.se}
\thanks{Present address: Department of Management Engineering, Technical University of Denmark, DTU, 2800 Kgs. Lyngby, Denmark}
\affiliation{$^1$Nordita, KTH Royal Institute of Technology and Stockholm University, Roslagstullsbacken 23, SE-106 91 Stockholm, Sweden\\
$^2$Department of Physics, Stockholm University, SE-10691 Stockholm, Sweden\\
$^3$Department of Computer Science, KTH Royal Institute of Technology, SE-10044 Stockholm, Sweden\\
$^4$Department of Physics, University of Connecticut, Storrs, CT 06269, USA
}
\author{Bart~Olsthoorn$^{1,2}$}
\affiliation{$^1$Nordita, KTH Royal Institute of Technology and Stockholm University, Roslagstullsbacken 23, SE-106 91 Stockholm, Sweden\\
$^2$Department of Physics, Stockholm University, SE-10691 Stockholm, Sweden\\
$^3$Department of Computer Science, KTH Royal Institute of Technology, SE-10044 Stockholm, Sweden\\
$^4$Department of Physics, University of Connecticut, Storrs, CT 06269, USA
}
\author{M.~Berk~Gedik$^{1,3}$}
\affiliation{$^1$Nordita, KTH Royal Institute of Technology and Stockholm University, Roslagstullsbacken 23, SE-106 91 Stockholm, Sweden\\
$^2$Department of Physics, Stockholm University, SE-10691 Stockholm, Sweden\\
$^3$Department of Computer Science, KTH Royal Institute of Technology, SE-10044 Stockholm, Sweden\\
$^4$Department of Physics, University of Connecticut, Storrs, CT 06269, USA
}
\author{R.~Matthias~Geilhufe$^{1}$}
\affiliation{$^1$Nordita, KTH Royal Institute of Technology and Stockholm University, Roslagstullsbacken 23, SE-106 91 Stockholm, Sweden\\
$^2$Department of Physics, Stockholm University, SE-10691 Stockholm, Sweden\\
$^3$Department of Computer Science, KTH Royal Institute of Technology, SE-10044 Stockholm, Sweden\\
$^4$Department of Physics, University of Connecticut, Storrs, CT 06269, USA
}
\author{Alexander~V.~Balatsky$^{1,4}$}
\affiliation{$^1$Nordita, KTH Royal Institute of Technology and Stockholm University, Roslagstullsbacken 23, SE-106 91 Stockholm, Sweden\\
$^2$Department of Physics, Stockholm University, SE-10691 Stockholm, Sweden\\
$^3$Department of Computer Science, KTH Royal Institute of Technology, SE-10044 Stockholm, Sweden\\
$^4$Department of Physics, University of Connecticut, Storrs, CT 06269, USA
}

\date{\today}

\begin{abstract}
We present an online graphical pattern search tool for electronic band structure data contained within the Organic Materials Database (OMDB) available at \url{https://omdb.diracmaterials.org/search/pattern}.
The tool is capable of finding user-specified graphical patterns in the collection of thousands of band structures from high-throughput \textit{ab initio} calculations in the online regime. Using this tool, it \change{only takes a few seconds} to find an arbitrary graphical pattern within the ten electronic bands near the Fermi level for \materialsN{} organic crystals. The tool can be used to find realizations of functional materials characterized by a specific pattern in their electronic structure, for example, Dirac materials, characterized by a linear crossing of bands; topological insulators, characterized by a ``Mexican hat'' pattern or an effectively free electron gas, characterized by a parabolic dispersion. \change{The source code of the developed tool is freely available at \url{https://github.com/OrganicMaterialsDatabase/EBS-search} and can be transferred to any other electronic band structure database.} The approach allows for an automatic online analysis of a large collection of band structures where the amount of data makes its manual inspection impracticable. 
\end{abstract}

\maketitle

\section{Introduction}\label{sec:intro}

Recent developments in materials informatics~\cite{rodgers2006materials,ferris2007materials} combined with ever-growing computational power have opened the way towards performing high-throughput calculations based on first-principles ({\it ab initio}) methods~\cite{10.1038/nmat3568}. This approach significantly facilitates the accelerated discovery of various materials with special functional properties~\cite{PhysRevX.1.021012,10.1038/nmat1752,doi:10.1021/cm200949v,klintenberg2014computational,geilhufe_point,geilhufe_line}. As a result, we witness an exponentially increasing amount of data usually organized in the form of databases like the Materials Project \cite{Rasmussen}, the Computational 2D Materials Database \cite{Jain2013} or the Organic Materials Database (OMDB) \cite{borysov2016}, to name but a few. To keep pace with the amount of data generated, there has to be a commensurate development of data mining and information retrieval tools capable of answering non-trivial questions about the data. Here, we present the first online graphical pattern search tool which is capable of finding user-specified graphical patterns in a collection of thousands of electronic band structures (EBS).

Recently, we witness an ongoing interest in extending the theory of electronic bands. This effort is mainly motivated by two ideas: the search for semimetals with low-energy excitations behaving as exotic quasi-particles \cite{bradlyn2016beyond} and the recent developments in the \change{topological band theory \cite{bradlyn2017topological,geilhufe_point,geilhufe_line,PhysRevB.94.155108,bouhon2017global,bzduvsek2016nodal}}. 
Realizations of non-trivial EBS features comprise the massless Dirac-fermions which were experimentally verified in graphene \cite{novoselov2005two} as well as the Weyl-fermions, which were found for instance in TaAs crystals \cite{Xu613}. With the introduction of the so-called Weyl type-II semimetals \cite{soluyanov2015type}---Weyl semimetals with heavily tilted energy-momentum cones---it is claimed that elementary excitations of the crystal can even mimic the physics of electrons close to the event horizon of black holes \cite{volovik2016lifshitz}. This interpretation suddenly opens the path to verify theoretical statements of black hole physics within relatively easily approachable measurements on single crystals. More exotic quasiparticles, which were discussed in a similar manner, are, for example, the double Dirac semimetal \cite{wieder2016double}, the node-line semimetals \cite{yu2015topological}, the hourglass fermions \cite{wang2016hourglass} or the triple-fermion materials \cite{lv2017observation}. 

To find real material realizations of these topological band features, manual inspection of EBSs 
represents a relatively easy task for a small number of materials. However, this approach becomes impracticable for thousands of band structures contained within modern EBS databases. Despite providing basic search functionality, most of the online databases lack non-trivial online search tools for EBS data querying and analysis. Our tool's software implementation based on the approximate nearest neighbor search algorithm is designed to match the constraints of web applications in terms of fast execution time and low memory usage. The tool is accessible within the web interface of the OMDB hosting thousands of EBSs for previously synthesized organic crystals at \url{https://omdb.diracmaterials.org/search/pattern}. \change{The source code of the developed tool is freely available at \url{https://github.com/OrganicMaterialsDatabase/EBS-search} and can be transferred to any other EBS database.}

The rest of the paper is organized as follows. In Results, we describe the pattern search tool interface and its implementation. In Discussion, application examples for the discovery of novel functional materials are shown.  Finally, technical details related to the OMBD data and pattern-matching algorithms are provided in Methods.

\section{Results}\label{sec:results}

\subsection{Pattern search algorithm}\label{sec:tool_back}

For a three-dimensional crystalline solid, the EBS is a four-dimensional object representing energy levels of electrons dependent on a three-dimensional momentum vector.
With aim to capture its most distinctive features in such cases, the EBS is usually calculated along specific paths within the Brillouin zone, for example, depending on the crystalline symmetry~\cite{SETYAWAN2010299}. Hence, properties of the EBS can be effectively characterized by one-dimensional patterns involving one or multiple bands. 

\begin{figure}[h]
\centering\includegraphics[width=1.0\linewidth]{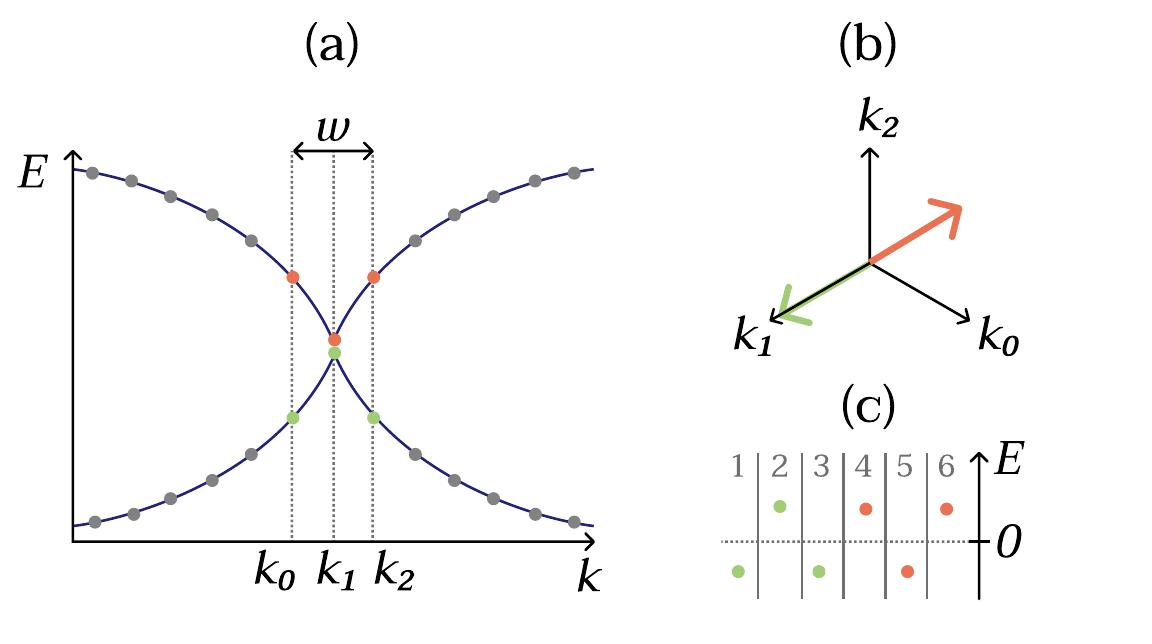}
\caption{A short summary of the pattern search algorithm. For each moving window of size $w$, $d$ points are selected from each band for the analysis. Although the dimension of an electronic band along some high-symmetry path in the Brillouin zone is one, the dimension of the corresponding feature space, being represented in a vector form, is defined by the number of points in it. For instance, for a moving window comprising 2 bands with 3 points each (a), the dimensionality of the corresponding feature space is 3 for each band (b) and 6 for the final concatenated vector (c). In the last step, the distance between the normalized concatenated vector and query pattern vector is calculated. 
}
\label{fig:principle}
\end{figure}

\change{To locate query patterns in the EBS data from the {\it ab initio} calculations stored in the OMDB}, we employ a moving window approach. Each continuous path in the Brillouin zone is scanned with a moving window of width $w$ in the momentum space with the stride $s$, specifying the number of data points the window jumps at each scanning step. \change{Since the EBS is calculated numerically along a discrete mesh with different spacing for different paths  within the Brillouin zone, linear interpolation is used to approximate energy values between the mesh points. For each moving window, we uniformly select $d$ energy values from each band and form a vector to be compared with a query pattern, being also represented as a vector in the same way (Fig.~\ref{fig:principle}a). Thus, in the case of a query pattern consisting of $n$ bands, the resulting vector dimensionality is $d \times n$ (Fig.~\ref{fig:principle}c). It is important to note that the present pattern search algorithm does not take into account the distance between bands (for instance, the distance between the maximum value of the lower band and the minimum value of the upper band in the $n=2$ case), which needs to be specified explicitly by the user.} 

To measure similarity between a vector obtained from the moving window and the query vector, the cosine distance $\sqrt{2-2\cos{\theta}}$ is used, where $\theta$ is the angle between \change{the normalized vectors. The normalization makes the cosine distance equivalent to the Euclidean ($L^2$) distance. It also makes the distance insensitive to energy scaling.} As $\theta$ ranges from $0$ (two vectors are the same) to $\pi$ (two vectors are opposite), the distance ranges from 0.0 to 2.0, respectively. \change{Finally, $K$ nearest vectors to the query vector are retrieved.} 

Unfortunately, finding the \change{nearest vectors becomes} computationally demanding with respect to memory and CPU usage, especially if it comes to applications within a web interface. \change{A straightforward exhaustive search algorithm, which goes through every vector, requires the number of comparisons equal to the total number of vectors to be queried.} For example, applying the moving window approach with the realistic parameters $w=0.4$, $d=16$ and $s=2$ for 10 bands near the Fermi surface for \materialsN{} materials within the OMDB produces \change{over $1.6\times 10^7$} vectors to query. As performance is crucial for online implementation, the exhaustive solution becomes impractical.

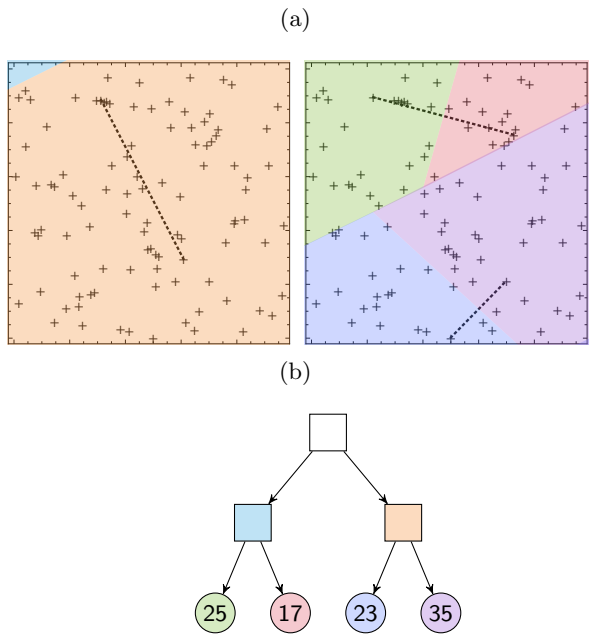
\begin{figure}[h!]
    \centering
    (a)\\
    \subfloat{\includegraphics[width=0.9\linewidth]{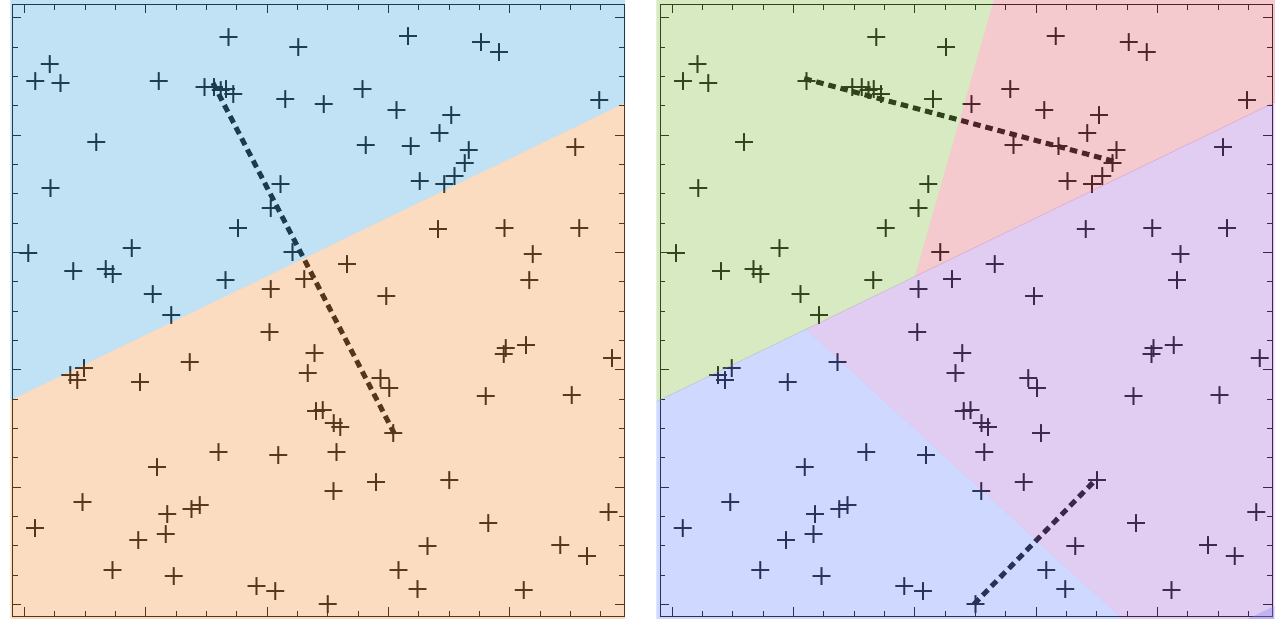}}\\
    (b)\\
    \subfloat{
\definecolor{blue}{rgb}{0.75,0.89,0.96}
\definecolor{orange}{rgb}{0.99,0.86,0.75}
\definecolor{purple}{rgb}{0.88,0.80,0.95}
\definecolor{green}{rgb}{0.84,0.92,0.76}
\definecolor{red}{rgb}{0.96,0.79,0.81}
\definecolor{blue2}{rgb}{0.81,0.85,1.0}
\tikzset{
    treenode/.style = {align=center, inner sep=0pt, text centered,
    font=\sffamily},
    arn_root/.style = {treenode, rectangle, black, draw=black,
    fill=white, text width=1.5em, minimum width=1.5em, minimum height=1.5em},
    arn_n1/.style = {treenode, rectangle, black, draw=black,
    fill=blue, text width=1.5em, minimum width=1.5em, minimum height=1.5em},
    arn_n2/.style = {treenode, rectangle, black, draw=black,
    fill=orange, text width=1.5em, minimum width=1.5em, minimum height=1.5em},
    arn_n3/.style = {treenode, circle, black, draw=black,
    fill=purple, text width=1.5em},
    arn_n4/.style = {treenode, circle, black, draw=black,
    fill=green, text width=1.5em},
    arn_n5/.style = {treenode, circle, black, draw=black,
    fill=red, text width=1.5em},
    arn_n6/.style = {treenode, circle, black, draw=black,
    fill=blue2, text width=1.5em},
    arn_r/.style = {treenode, circle, red, draw=red, 
    text width=1.5em, very thick},
    arn_x/.style = {treenode, rectangle, draw=black,
    minimum width=0.5em, minimum height=0.5em}
}
\begin{tikzpicture}[->,>=stealth',level/.style={sibling distance = 2cm/#1,
  level distance = 1.2cm}, node distance= 2cm] 
    \node [arn_root] {}
        child { node [ arn_n1] {\ }
            child { node [ arn_n4] {25} }
            child { node [ arn_n5] {17} }
        }
        child { node [ arn_n2] {\ }
            child { node [ arn_n6] {23} }
            child { node [ arn_n3] {35} }
        };
        \node[] at (0cm, -3cm) {};
\end{tikzpicture}}

\caption{\change{An example of the ANNOY algorithm for 100 points in a 2D space. (a) First, the space is split in two subspaces. The split occurs as the equidistant hyperplane between two randomly selected points indicated by the dashed line. For each subspace, this step is repeated recursively, until the number of points in a subspace is below a certain threshold. (b) The constructed binary tree allows to find nearest neighbors in logarithmic time. The algorithm generalizes to higher dimensional spaces. For instance, for a pattern consisting of 2 bands with 3 points each, the dimensionality of the corresponding search space is 6.}}
    \label{fig:annoy_scheme}
\end{figure}

The exhaustive search can be accelerated with a computation-memory trade-off using a precalculated index structure based on search space partitioning. We implemented fast data access using the \change{open-source} ANNOY library~\cite{annoy}, which uses the approximate nearest neighbor search algorithm. During the indexing step, it creates multiple binary tree structures, where each intermediate node represents a split and each leaf node represents an area in the search space \change{(Fig.~\ref{fig:annoy_scheme})}. This precalculated index allows to significantly reduce the search time. More details about the approximate nearest neighbor algorithm can be found in Methods.

Since the bands near the Fermi level are usually of physical interest, we have indexed the 9 closest pairs of bands (5 bands above and 5 below the Fermi level). Thus, at the current stage, only these bands are available for the online search. We started with the implementation for the patterns consisting of two bands. However, the approach can be extended in a similar manner to patterns involving an arbitrary number of bands.

\subsection{The tool's interface}\label{sec:tool_front}

\begin{figure}[b] 
\centering\includegraphics[width=0.82\linewidth]{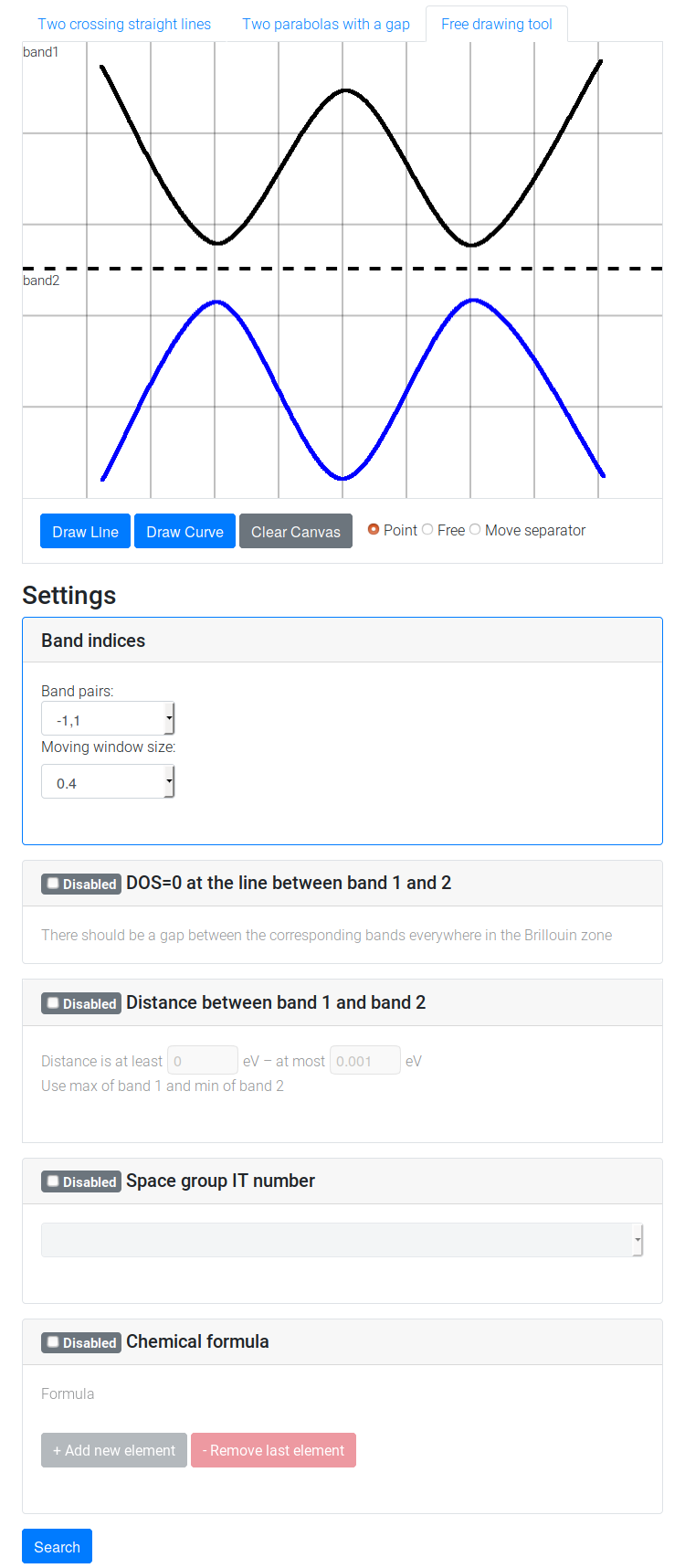}
\caption{The web interface of the pattern search tool. A user can either select a predefined pattern or use a free drawing input interface to search for an arbitrary pattern (a sketch of ``Mexican hat'' is shown). Also, a user can specify bands of interest, moving window size, distance and density of states between the bands in the pattern, along with other basic filtering options like space group number or chemical composition of the materials of interest.}
\label{fig:interface}
\end{figure}

The developed pattern search tool is available online at \url{https://omdb.diracmaterials.org/search/pattern}. The tool's web interface is shown in Fig.~\ref{fig:interface}. A user can either select one of the predefined query patterns (two crossing straight lines or two parabolas) or use a free drawing input interface to search for an arbitrary pattern. \change{Also, a user can specify the band indices with respect to the Fermi level where the search is performed, the moving window size in the momentum space, the maximum/minimum distance between the bands, if zero density of states between the bands is required, and other basic filtering options, such as space group number or chemical composition of the materials of interest.}

\subsection{Performance tests and calibration}
To test and calibrate our tool, we use the EBS data contained within the OMDB. \change{We also provide additional synthetic data tests together with the source code at \url{https://github.com/OrganicMaterialsDatabase/EBS-search}.}

\begin{figure}
\centering
(a)\hspace{3.6cm}(b)\\
\includegraphics[trim={1.0cm 1.0cm 1.0cm 1.0cm},clip,width=0.45\linewidth]{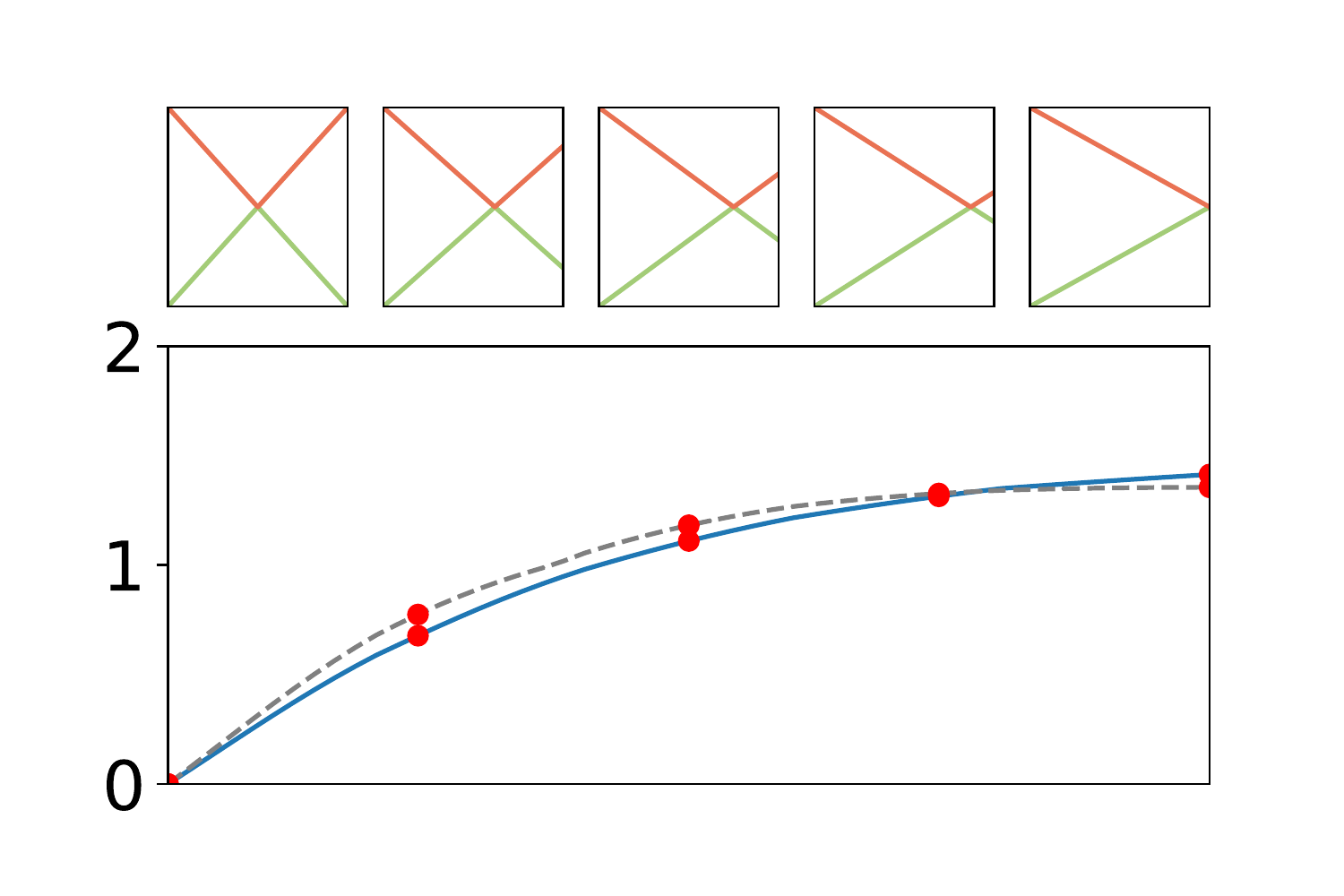}
\includegraphics[trim={1.0cm 1.0cm 1.0cm 1.0cm},clip,width=0.45\linewidth]{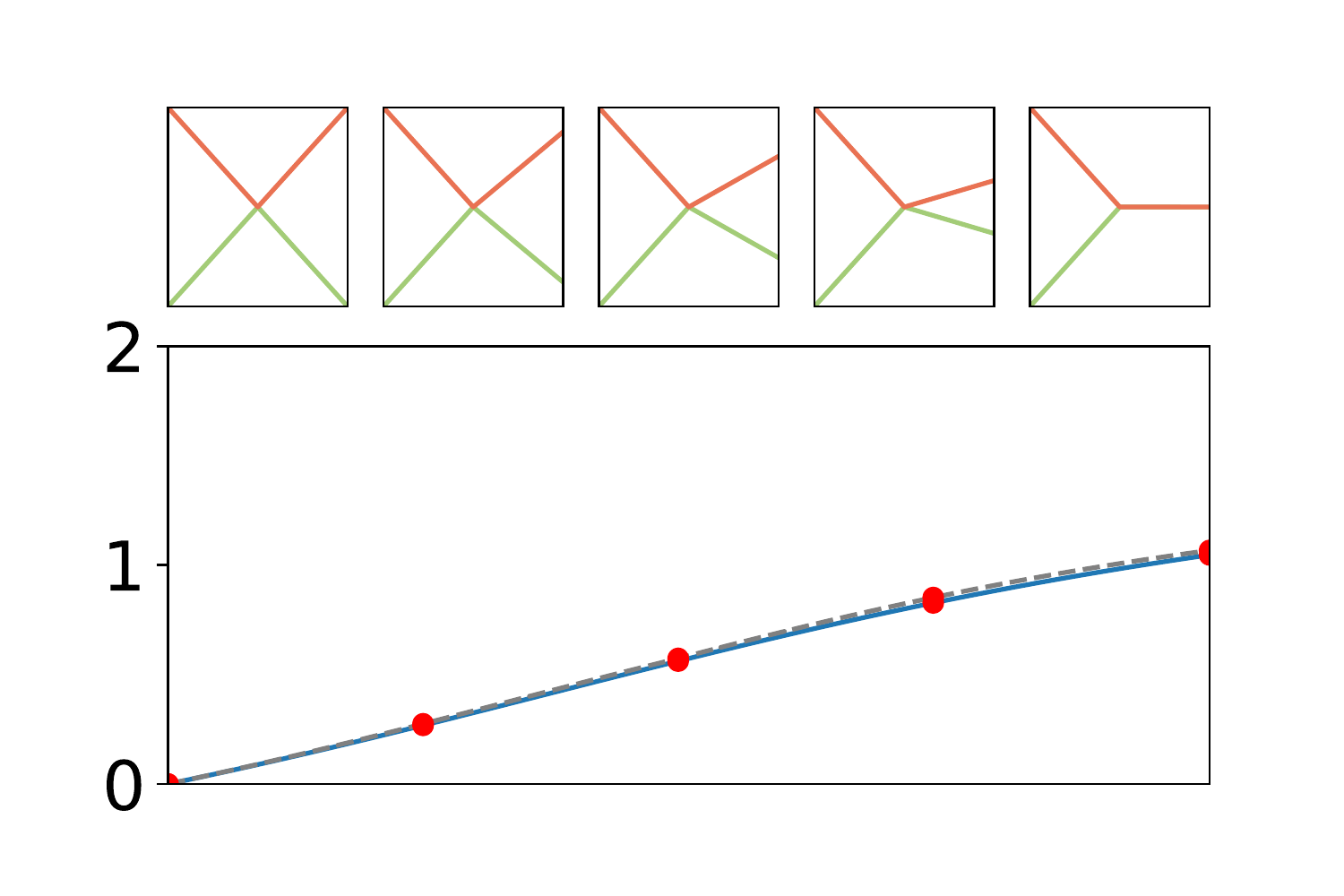}\\
(c)\hspace{3.6cm}(d)\\
\includegraphics[trim={1.0cm 1.0cm 1.0cm 1.0cm},clip,width=0.45\linewidth]{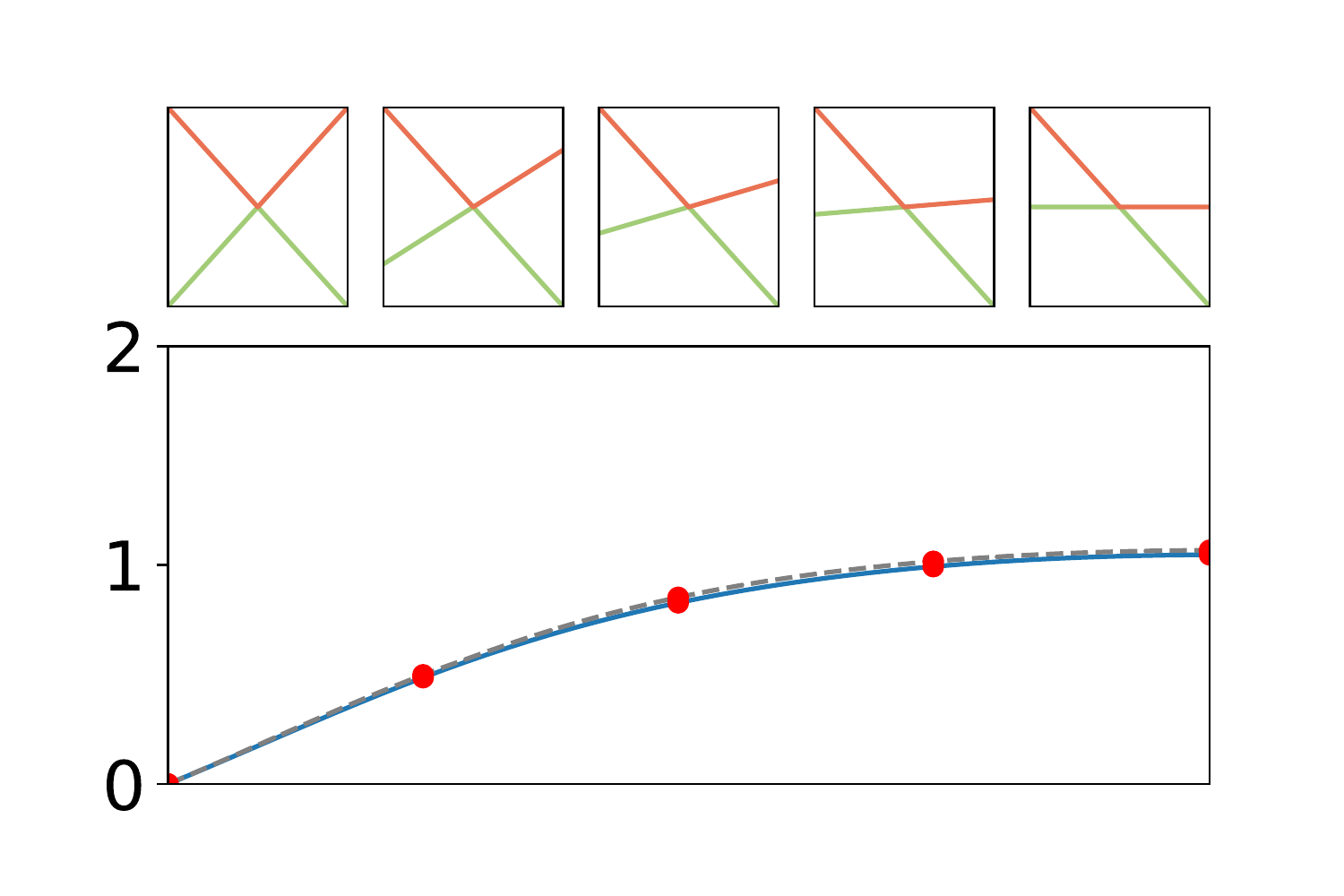}
\includegraphics[trim={1.0cm 1.0cm 1.0cm 1.0cm},clip,width=0.45\linewidth]{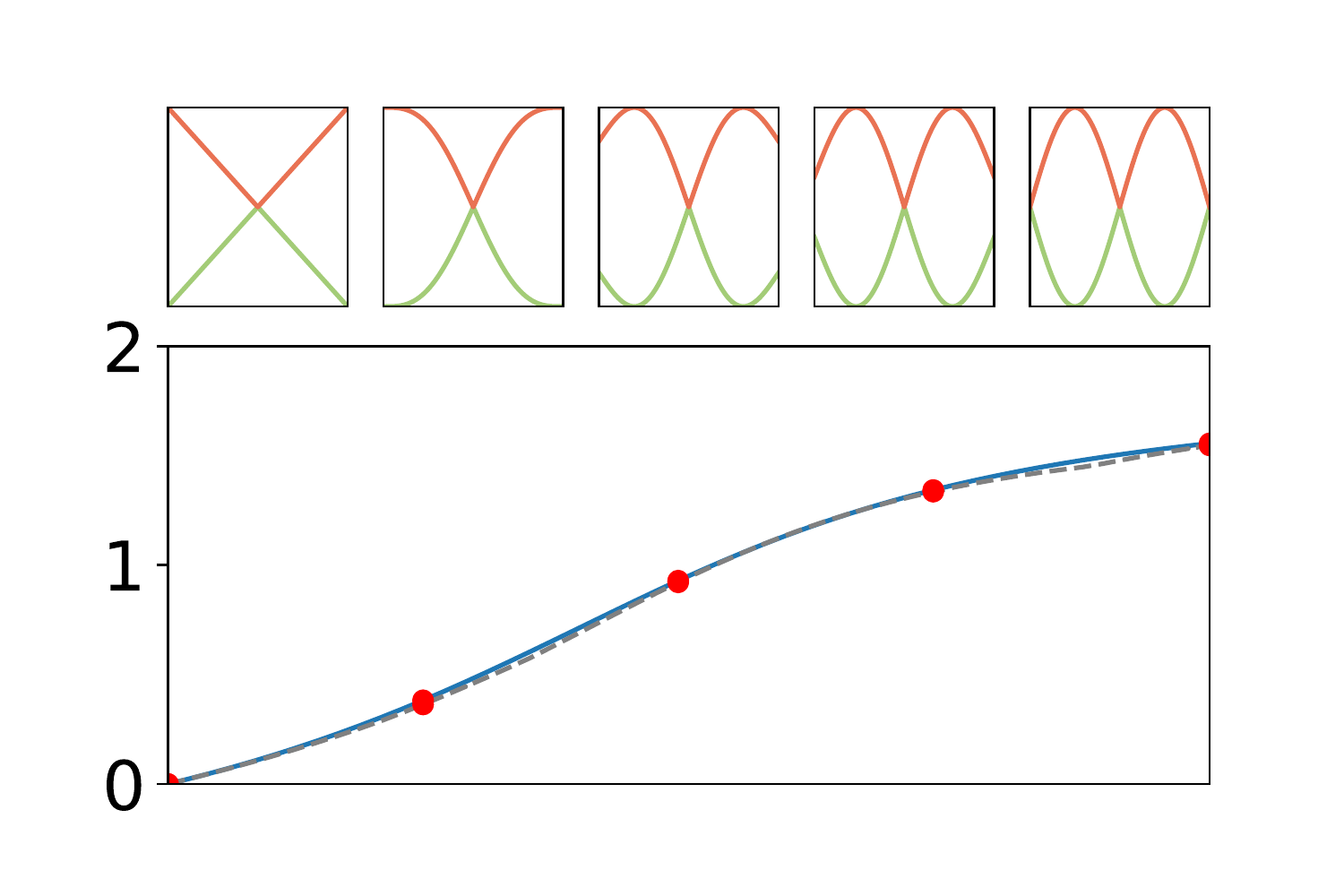}
\caption{\change{Sensitivity of the cosine ($L^2$) distance (solid blue line) and scaled Manhattan ($L^1$) distance (dashed gray line) to various distortions of the Dirac crossing pattern: (a) shift, (b) oblique, (c) skew and (d) nonlinear distortion/change of the characteristic scale. The distorted patterns are shown for the red dots. High-frequency noise and outliers are not included because band structures are usually smooth objects with low variance over a characteristic scale.}}
\label{fig:sens}
\end{figure}

\begin{figure}[h!]
\centering
(a)\\
\includegraphics[trim={0.5cm 16.6cm 0 0},clip,width=1.0\linewidth]{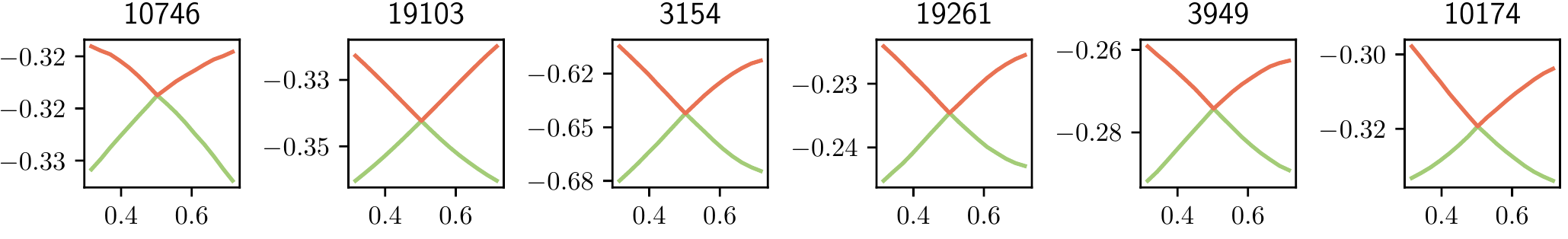}\\
\includegraphics[trim={0.5cm 0 0 0},clip,width=1.0\linewidth]{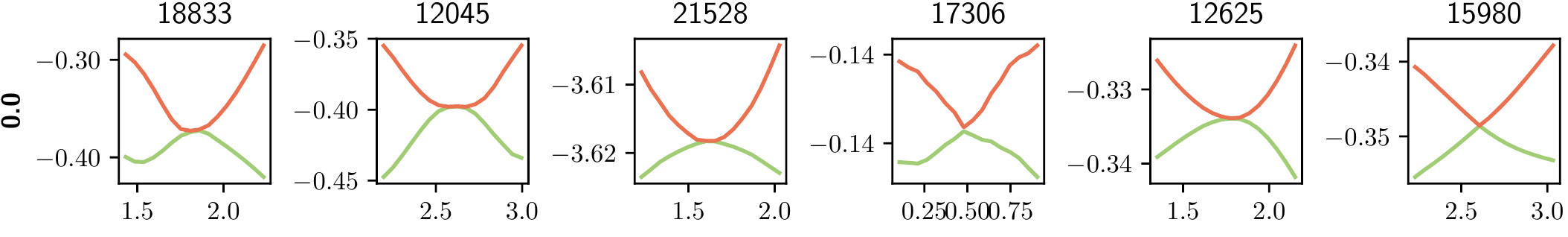}\\
(b)\\
\includegraphics[trim={0.6cm 16.6cm 0 0},clip,width=1.0\linewidth]{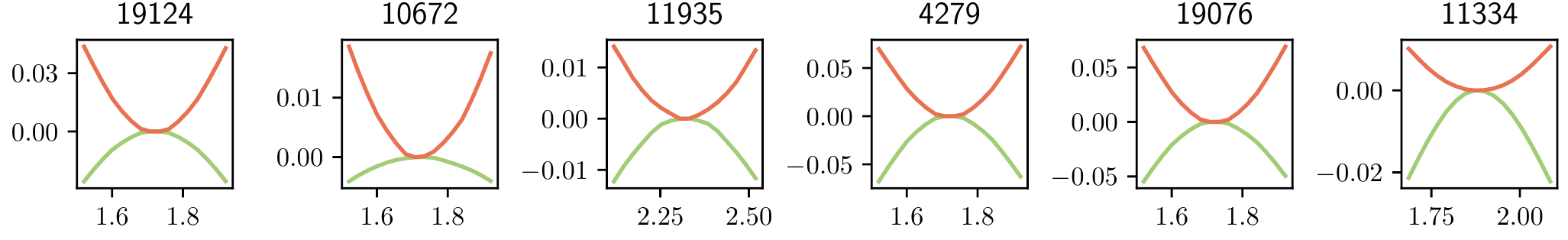}
\includegraphics[trim={0.5cm 0 0 0},clip,width=1.0\linewidth]{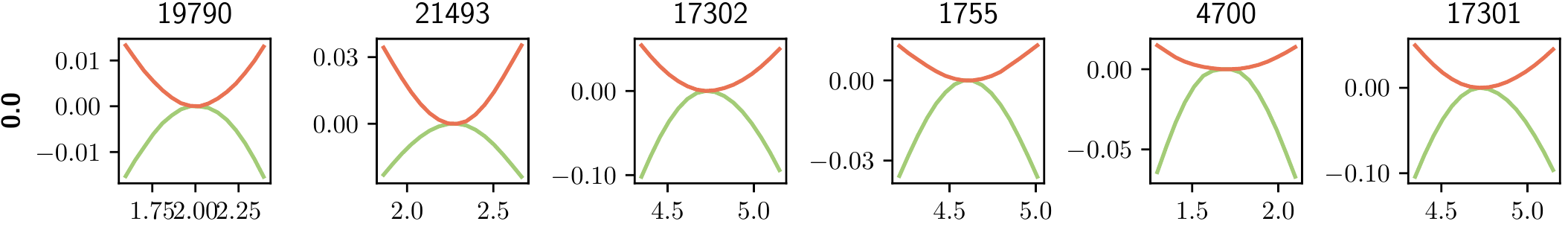}
\caption{Comparison of the top 6 search results for linear crossings (a) and two gapped parabolas (gap is not shown) (b) for two different moving window sizes: $0.4$ (first row) and $0.8$ (second row). The top search results for the linear crossings have much better quality for $w=0.4$ than for $w=0.8$, while the search for two gapped parabolas gives qualitatively acceptable results for both moving window sizes. The titles above the graphs indicate the OMDB-ID. The values for $E$ and $k$ match the values on the website.}
\label{fig:inspection_window08}
\end{figure}

\change{The first parameters to be defined are the moving window size $w$ and the stride $s$. With this aim, we test the sensitivity of the cosine distance to the various distortions of the search pattern. The results are shown in Fig.~\ref{fig:sens}. As can be verified, the distance between the query pattern and the example increases introducing shifts, obliques, skews or other nonlinear distortions.}
While $s$ should be small with respect to $w$ not to miss any possible search results (we use $s=2$ DFT mesh points), the moving window size $w$ is more task-specific. It should correspond to the expected characteristic momentum scale of the pattern of interest. For example, Fig.~\ref{fig:inspection_window08}a suggests that the top search results for a linear crossing pattern show a much better agreement for a window size of $w=0.4$ than for $w=0.8$. At the same time, a similar test for two gapped parabolas gives qualitatively acceptable results for both moving window sizes (Fig.~\ref{fig:inspection_window08}b). As $w$ is pattern-dependent, its value should be specified by the user. Furthermore, it is worth noting that for smaller values of $w$, we are restricted by the mesh resolution in the momentum space \change{stemming} from the {\it ab initio} calculations. For example, for the EBSs contained within the OMDB, the moving window for $w=0.4$ contains only 14.4 mesh points per band on average (minimum 9 and maximum 33). 

\begin{figure}[h!]
\centering
(a)\\
\includegraphics[width=0.98\linewidth]{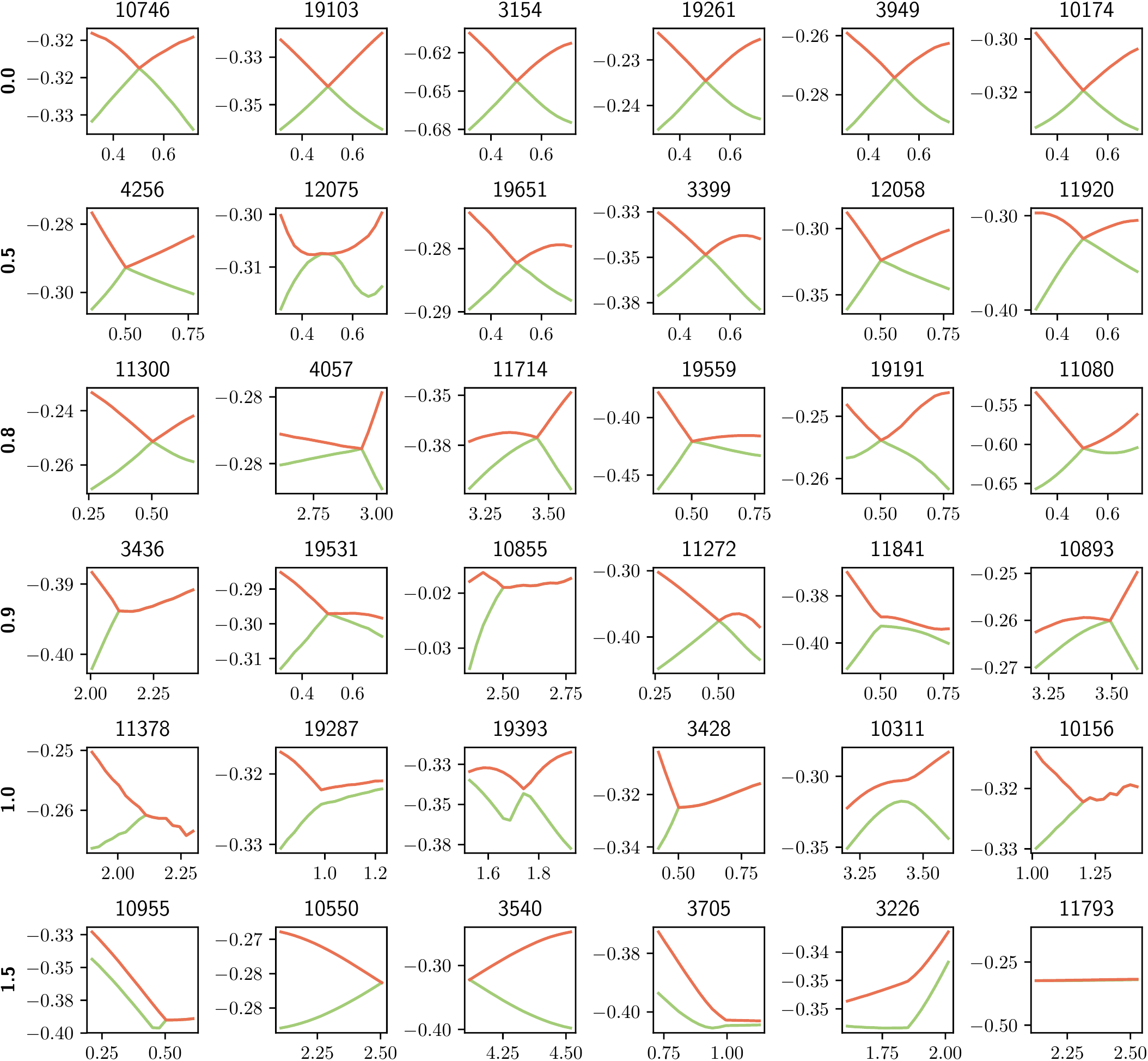}\\
(b)\\
\includegraphics[width=0.98\linewidth]{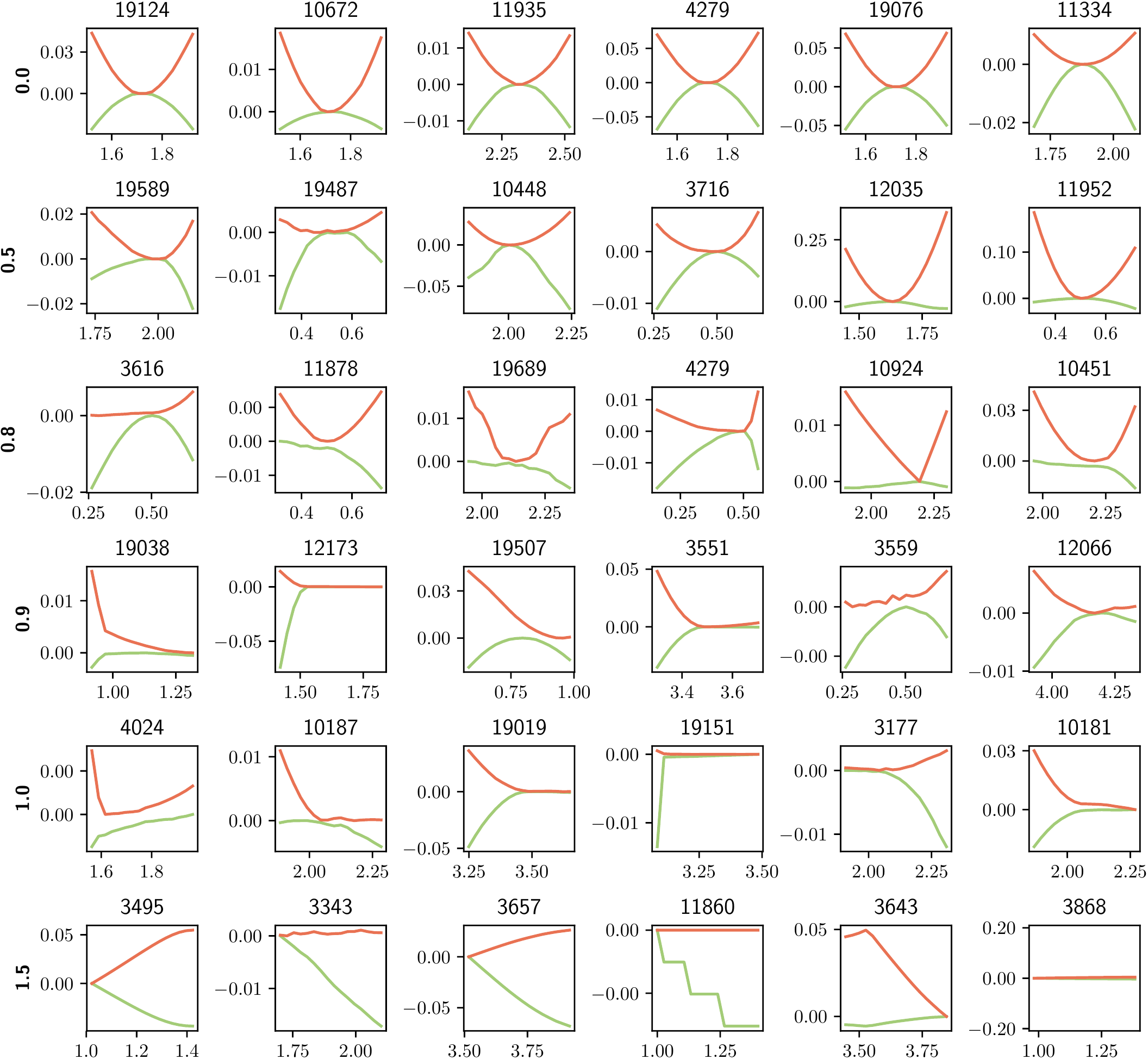}
\caption{Pattern search results for a linear crossing in the two highest valence bands (a) and two parabolas in the highest valence and lowest conduction bands (b). Each row shows the nearest vectors (best search results) starting from a distance threshold, for threshold values 0.0, 0.5, 0.8, 0.9, 1.0 and 1.5, respectively, for the moving window size of 0.4. The distance between upper and lower bands was set to be less than $0.0001$~eV for (a) and was not restricted for (b). The titles above the graphs indicate the OMDB-ID. The values for $E$ and $k$ match the values on the website.
}
\label{fig:inspection}
\end{figure}

It is also important to check a maximum value of the distance for a search result to be of acceptable quality. Since similarity to a pattern is an essentially subjective quality specific to the task in hand, we resort to visual inspection of the search results. Fig.~\ref{fig:inspection} shows that this value can vary from 0.8 for a linear crossing (Fig.~\ref{fig:inspection}a) to 0.5 for two gapped parabolas (Fig.~\ref{fig:inspection}b). On the website, we show the top search results ranked by their distance to the query pattern and use this threshold value in a warning message only.

As mentioned before, the exact nearest neighbor search algorithm is not applicable in the context of a web application due to the high computational demand. To tackle this issue, we choose the approximate nearest neighbor algorithm implemented in the ANNOY library, which has two parameters to tune: the number of search trees, $N$, and number of points to examine, $K$. Increasing both parameters gives more precise search results in expense of computational resources. Namely, $N$ affects the memory usage and $K$ affects the search time. 

\begin{figure}[h]
\centering
(a)\hspace{4cm}(b)\\
\includegraphics[width=0.46\linewidth]{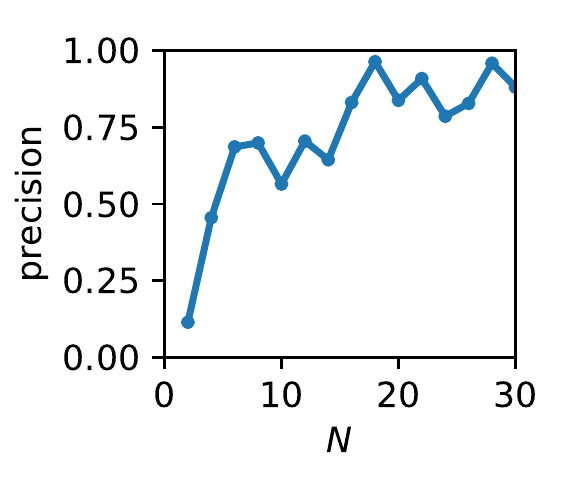}
\includegraphics[width=0.5\linewidth]{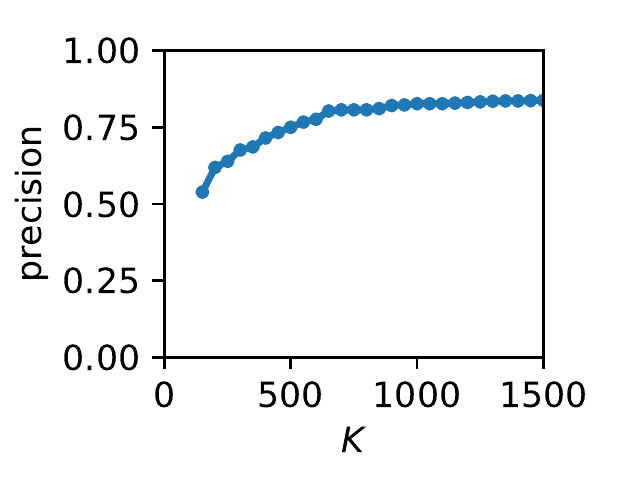}
\caption{The quality of the top 100 search results obtained using the ANNOY library grows with the number of trees $N$ for fixed $K=1500$ (a) and the number of leaf nodes $K$ for fixed $N=20$ (b). As a ground truth, we used the top 100 search results from the exact algorithm for the linear crossing pattern with a moving window size of $w=0.4$ in the two highest valence bands. The precision is calculated as the fraction of coinciding search results and micro-averaged over 10 different ANNOY indices.}
\label{fig:approx}
\end{figure}

To estimate these parameters, we compare the performance of the top 100 search results of the approximate nearest neighbor search algorithm to those of the exact algorithm for different values of $N$ and $K$. As a ground truth, we use the top 100 exhaustive search results with $w=0.4$ for the linear crossing pattern in the two bands below the Fermi level. As can be seen in Fig.~\ref{fig:approx}, the performance of the approximate nearest neighbor search is close to the exact solution, but allows to significantly reduce the search time. For example, using the values $N=20$ and $K=1500$, the approximate search is more than two orders of magnitude faster in comparison to the exact algorithm by obtaining comparable search results. The level of approximation can be always adjusted to the computational resources available.

\section{Discussion}\label{sec:discussion}

\change{It has been shown by several research groups that} the data mining approach has been successful, for example, for the search of stable nitride perovskites~\cite{sarmiento2015prediction}, thermoelectric materials \cite{PhysRevX.1.021012}, 
electrocatalytic materials for hydrogen evolution \cite{10.1038/nmat1752} or lithium-ion battery cathodes \cite{doi:10.1021/cm200949v}. Using a pattern search analysis of the data within the Electronic Structure Project \cite{ortiz2009data}, Klintenberg \textit{et al.} identified 17 candidates for strong topological insulators by mining for materials exhibiting the specific ``Mexican hat'' shaped dispersion relation \cite{klintenberg2014computational}. Similarly, by searching for linear crossings in band structures, novel Dirac materials can be identified as recently shown using the data within the OMDB \cite{geilhufe_point,geilhufe_line} and the Materials Project database \cite{yan2017data}. Alternatively, new functional materials can be predicted by comparison of specific features within the EBSs of known prototype materials to the EBSs within electronic structure databases, as shown for example in the case of potential high-temperature superconductors \cite{klintenberg2013possible,geilhufe2017novel}. Similar statistical methods can be also used to identify systematic trends in strongly correlated $f$-electron materials \cite{PhysRevMaterials.1.033802}. 

Here, we present a new approach to search for novel functional materials characterized by a specific pattern in their electronic structure, \change{such as Dirac materials, topological insulators, and novel semimetals with low-energy excitations behaving as exotic quasi-particles.}

\begin{figure}[b]
\centering
(a)\\
\includegraphics[height=4.1cm]{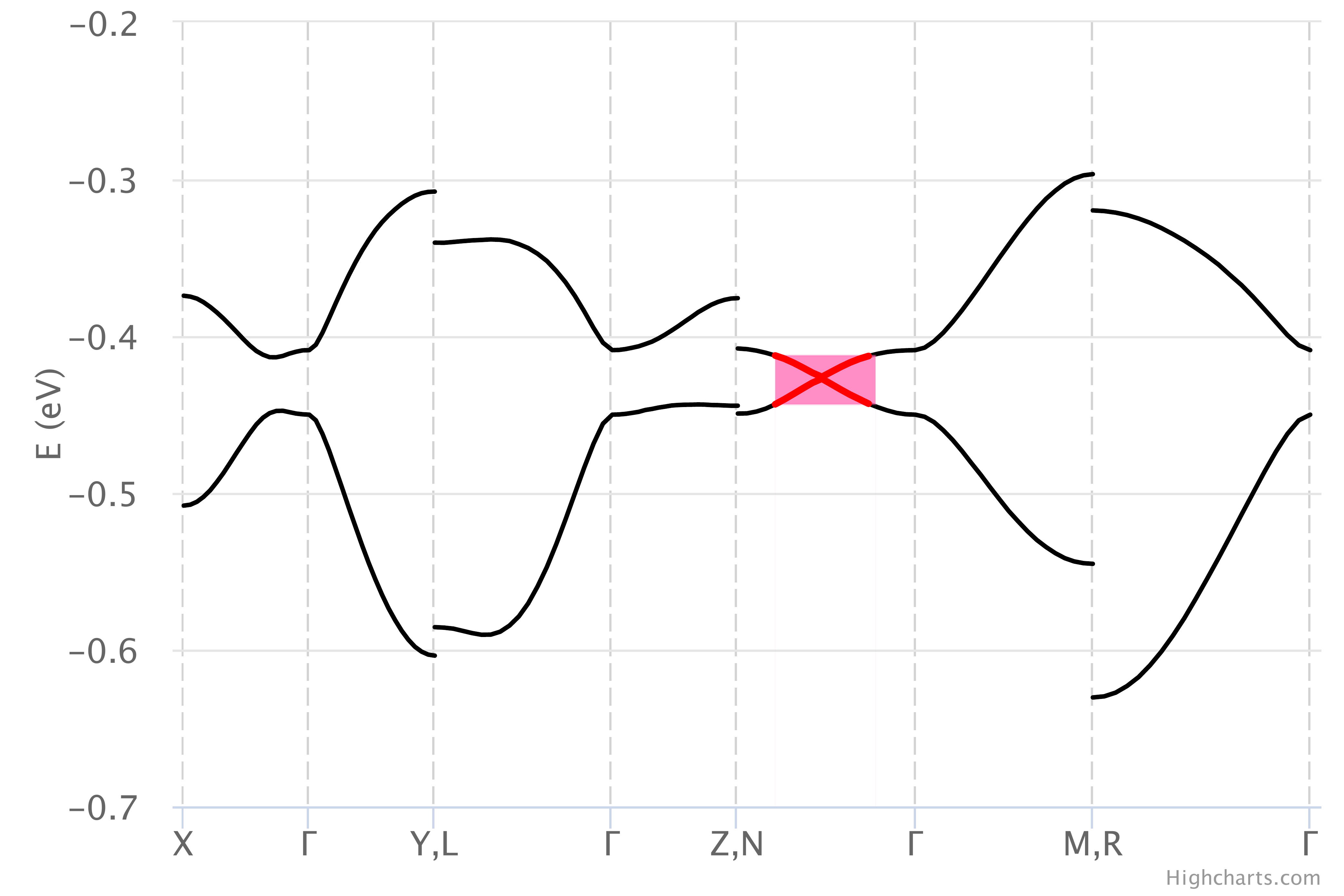}\\
(b)\\
\includegraphics[height=4.1cm]{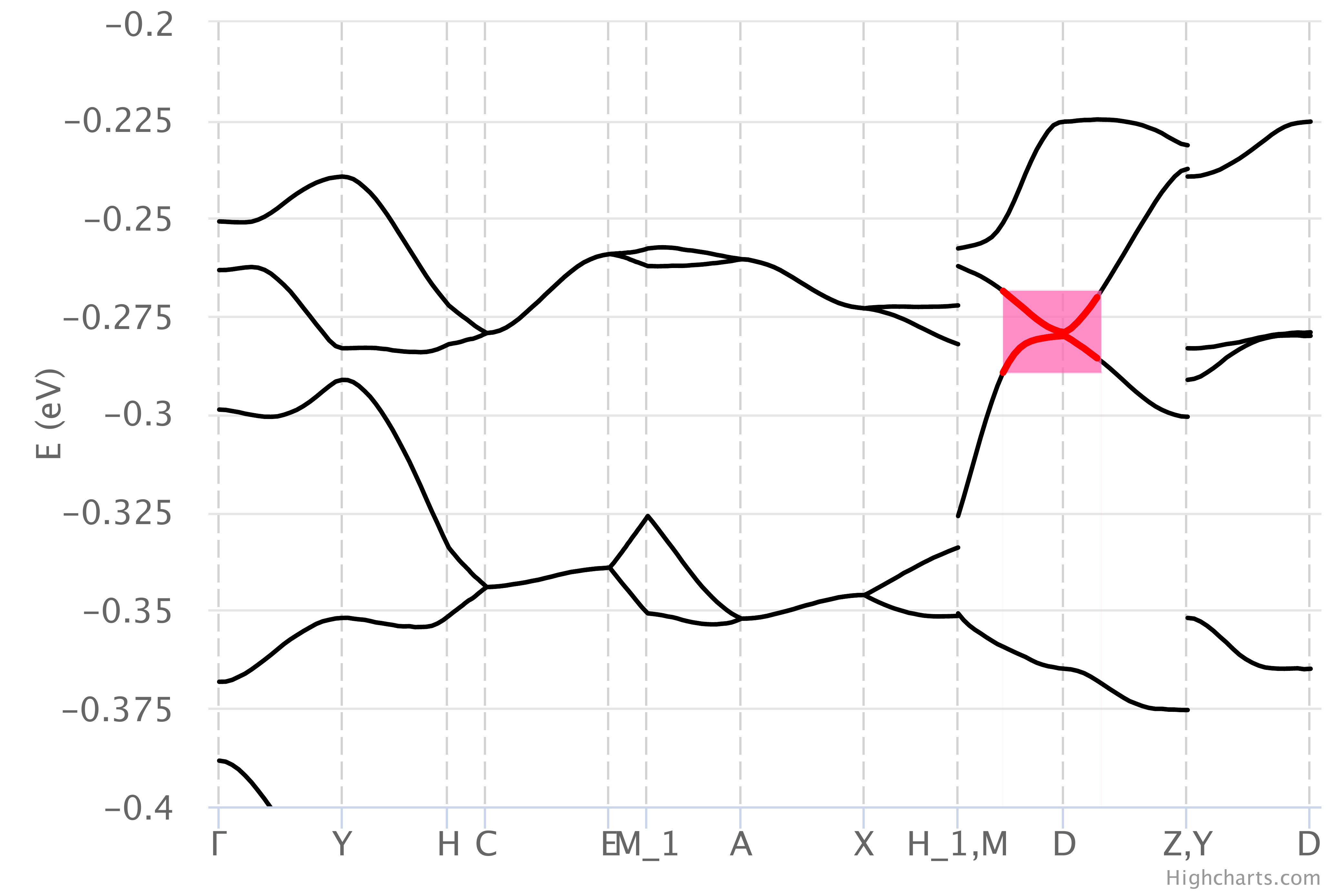}\\
(c)\\
\includegraphics[height=4.1cm]{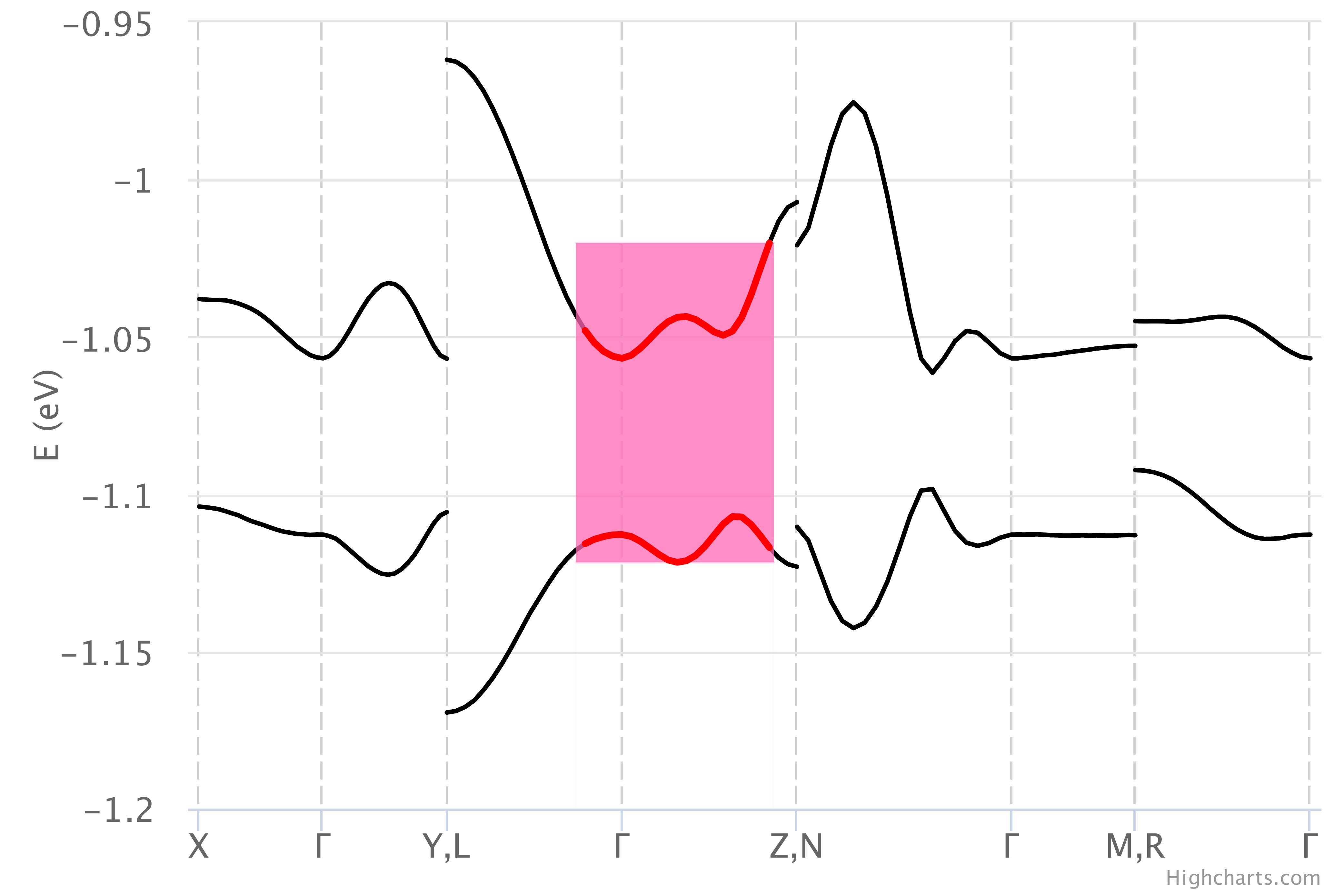}
\caption{Examples of search results for the patterns which might be interesting from a physical point of view:  Dirac crossing, OMDB-ID 4381 (a); two touching parabolas, OMDB-ID 4492 (b); Mexican hat, OMDB-ID 2308 (c). \change{Plotted using Highcharts library~\cite{highcharts_url}.}}
\label{fig:results}
\end{figure}

A data-mining approach by means of the described pattern-matching algorithm can be a powerful tool. As the first example, we consider the linear crossing of two bands indicating Dirac materials. This class of materials has been extensively studied due to the exceptional transport and optical properties~\cite{RevModPhys.83.407,abergel2010properties}. 
To achieve an isolated crossing in the energy space, the additional constraint of having vanishing density of states at the crossing point was applied. Since the majority of organic crystals are insulating \cite{borysov2016}, we searched for the pattern in the first and second highest valence bands. The maximum band distance was set to 0.01~eV and the moving window size was restricted to 0.4. Using this conditions, the algorithm found \change{51} matching results, where the best one has the match error of 0.075 and band distance of 0~eV. The corresponding band structure is plotted in Fig.~\ref{fig:results}a, which belongs to the material C$_9$H$_5$ClN$_2$O$_2$ (OMDB-ID 4381, COD-ID 7155013), crystallizing in a triclinic crystal. It is also worth mentioning that, using an offline version of the presented tool, several novel organic Dirac materials have been already predicted~\cite{geilhufe_line,geilhufe_point}.

Whereas a linear crossing of bands corresponds to a nearly free electron gas of massless Dirac fermions, two touching parabolas mimic the behavior of massive free electrons corresponding to the Schr\"odinger equation. However, the search for two touching parabolas did not retrieve any materials with vanishing density of states at the touching point. Having weakened this criteria, the search for two touching parabolas within the second and third valence bands retrieved 1443 materials with the matching error for the top result of 0.224. The corresponding band structure is illustrated in Fig.~\ref{fig:results}b, which belongs to C$_{20}$H$_{20}$BrN$_3$O$_3$ (OMDB-ID 4492, COD-ID 7153203), having a monoclinic crystal structure.

Next to semimetals, materials possessing a gap can also show specific patterns. The most relevant examples are the topological insulators \cite{hasan2010colloquium}, where an overlap of two bands combined with a forbidden crossing leads to the specific Mexican hat shape of bands. This phenomenon is also referred to as band inversion. While the bulk of a topological insulator is insulating, metallic states on the surface can be found as a consequence of the topological gap. Well known examples comprise the materials Pb$_x$Sn$_{1-x}$Te~\cite{tanaka2012experimental,geilhufe2015effect,hsieh2012topological} or Bi$_2$Se$_3$~\cite{chen2009experimental}. The theory of topological gaps is clearly not restricted to a band gap at the Fermi level, but can be generalized to any occurring spectral gap within the band structure. By searching for the Mexican hat shape in the third and forth bands below the Fermi level, we found 290 materials within a moving window size of 0.8. The band distance was allowed to be in the range of 0.05~eV to 9~eV and the density of states was forced to be zero between the bands. As an example, the material C$_{11}$H$_{17}$ClO$_2$ (OMDB-ID 2308, COD-ID 4030217) was found with the match error of 0.59 (Fig.~\ref{fig:results}c).

\section{Methods}\label{sec:mm}
\subsection{Organic Materials Database (OMDB)}\label{sec:omdb}
The Organic Materials Database (OMDB) \cite{borysov2016} is an online database available at  \url{https://omdb.diracmaterials.org} containing the output of \textit{ab initio} calculations based on density functional theory (DFT) \cite{Hohenberg1964,Kohn1965} for \materialsN{} (at the moment of writing) previously synthesized three-dimensional organic crystal structures taken from the Crystallography Open Database (COD) \cite{Grazulis2009}. The DFT calculations were performed using the Vienna Ab initio Simulation Package (VASP)~\cite{vasp3}. The OMDB contains EBSs calculated along high symmetry $\vec{k}$-paths in the Brillouin zone which were automatically generated by the Pymatgen package \cite{Ong2013314}. Electronic bands for each path were calculated on a discrete mesh consisting of 20 points independently of its length in the momentum space. 
For the pattern search, we use continuous paths suggested by Pymatgen. However, we plan to extend the search to 
cover all possible combinations of calculated paths sharing the same high-symmetry point. Although the calculations were performed spin-polarized, we do not distinguish between spin-up and spin-down bands for the pattern search task. More details about the DFT calculations can be found in Ref.~\cite{borysov2016}. 

\change{
\subsection{Problem overview}\label{sec:problem}
The problem of locating patterns similar to a target (query) pattern in a sequence of data points has a long interdisciplinary history. Related approaches are typically based on scanning the sequence with a moving window followed by comparison of these shorter subsequences with the query~\cite{abdula1}. This approach has several dimensions to explore. The first one is related to the data representation. As an alternative to the raw data points, a fitted model or a transformation, such as Fourier \cite{10.1007/3-540-57301-1_5}, wavelet \cite{abdula6} or dimensionality reduction \cite{abdula23}, can be employed. Second, a similarity measure between the subsequences and the query needs to be defined. Most of them are based on the $L^p$-norms, however, more advanced probability measures~\cite{Keogh} have also been discussed. Finally, for practical applications, an efficient search algorithm is necessary. Usually, it involves indexing the subsequences obtained by a moving window with a tree-like partition structure.
The presented solution in this paper uses a cosine similarity (equivalent to the $L^2$ distance for normalized vectors) and binary search trees as implemented in the open-source ANNOY library~\cite{annoy}. No advanced data transformations are used.}

\subsection{Nearest neighbor search algorithm}\label{sec:pattern_ebs}

\change{The main idea of the nearest neighbor search~\cite{yianilos1993data} is to find nearest vectors to a query vector given some distance measure.}
The most straightforward (exact) nearest neighbor algorithm \change{iterates through each vector and calculates the distance to the query.}
This \change{linear complexity} algorithm can be accelerated with a computation-memory trade-off using a pre-calculated index structure based on search space partitioning. However, the related algorithms are not exact anymore, because they can miss some search results. Nevertheless, due to the high computational demand of the exact search, it becomes necessary to use an approach which returns ``close enough'' neighbors in order to obtain a good speed improvement. In many cases, approximate methods perform comparably to the exact one \cite{4031381}. Many open-source libraries are available where various indexing strategies and approximation methods have been implemented, for example, ``FAISS'' released by Facebook AI Research \cite{DBLP:journals/corr/JohnsonDJ17}, ``ANNOY'' by Spotify \cite{annoy} and Non-Metric Space Library (NMSLIB) \cite{DBLP:conf/sisap/BoytsovN13}.

The back-end of the graphical pattern search tool is implemented using the \change{open-source} ANNOY library \cite{annoy} which is based on the approximate nearest neighbor search. During the indexing step, it creates a binary tree structure for the data \change{vectors} where each intermediate node represents a split and each leaf node represents an area in the search space. It keeps splitting the space randomly using equidistant hyperplanes between two randomly selected \change{vectors} in each node until the number of \change{vectors} in a subspace is below a certain threshold. It can also use multiple trees $N$ (\verb|n_trees| in the ANNOY documentation) in order to improve the quality of search results in expense of memory usage. When a user tries to find closest neighbors of a query vector, the library first finds the leaf node that the query vector would belong to and collects $K$ \change{vectors} to test (\verb|search_k| in the {ANNOY} documentation) from that node as well as nearby \change{leaf} nodes. Then, it eliminates duplicates, which come from different trees, and calculates the distance between each selected \change{vector} and the query. Here, $N$ and $K$ can be tuned to find a trade-off between the algorithm's precision and performance.

\section*{Acknowledgments}
We are grateful for support from the Villum Foundation, Swedish Research Council Grant No.~638-2013-9243, the Knut and Alice Wallenberg Foundation and the European Research Council under the European Union’s Seventh Framework Program (FP/2207-2013)/ERC Grant Agreement No.~DM-321031. The authors acknowledge computational resources from the Swedish National Infrastructure for Computing (SNIC) at the National Supercomputer Centre at Link\"oping University as well as the High Performance Computing Center North.

\bibliography{references.bib}

\end{document}